\documentstyle[11pt,paspconf,psfig]{article}

\def\DF{DF}
\def\eg{{\it e.g.}}
\def\etal{{\it et al.}}
\def\etc{{\it etc.}}
\def\ie{{\it i.e.}}
\def\spose#1{\hbox to 0pt{#1\hss}}
\def\ltsim{\mathrel{\spose{\lower.5ex\hbox{$\mathchar"218$}}
     \raise.4ex\hbox{$\mathchar"13C$}}}
\def\gtsim{\mathrel{\spose{\lower.5ex \hbox{$\mathchar"218$}}
     \raise.4ex\hbox{$\mathchar"13E$}}}

\begin{document}
\title{Spiral Structure as a Recurrent 
Instability\altaffilmark{1}\altaffilmark{2}}

\author{J. A. Sellwood}
\altaffiltext{1}{Rutgers Astrophysics Preprint 256}
\altaffiltext{2}{To appear in {\it Astrophysical Dynamics -- in 
commemoration of F. D. Kahn}, eds.\ D. Berry, D. Breitschwerdt, A. 
da Costa \& J. E. Dyson (Dordrecht: Kluwer)}

\affil{Department of Physics \& Astronomy, Rutgers University, \\
136 Frelinghuysen Road, Piscataway, NJ 08854, USA\\
email: sellwood@astro.rutgers.edu}

\begin{abstract}
A long-standing controversy in studies of spiral structure has 
concerned the lifetimes of individual spiral patterns.  Much 
theoretical work has sought quasi-stationary spiral modes while 
$N$-body simulations have consistently displayed recurrent, 
short-lived patterns.  The simulations manifest a recurrent cycle of 
true instabilities related to small-scale features in the angular 
momentum distribution of particles, with the decay of each 
instability seeding the growth of the next.  Data from the recent 
Hipparcos mission seem to offer support for the recurrent transient 
picture.
\end{abstract}
\keywords{}

\section{Introduction}
It is about a century and a half since Lord Rosse first noted the 
spiral appearance of M51, but a satisfying and robust theory for the 
general spiral phenomenon in disc galaxies still eludes us.

There has been substantial progress, of course.  Early theoretical 
efforts focused on the gas, no spirals are seen in S0 galaxies that 
have little or no interstellar matter, and are most striking where 
gas is abundant.  However, most researchers are now convinced that 
spirals are driven by the stellar disc through some kind of 
collective gravitational process.  The most compelling reason is 
that spiral arms are smoother in images of galaxies in the near IR 
(Schweizer 1976; Rix \& Zaritsky 1995; Block \& Puerari 1999), 
indicating that the old disc stars participate in the pattern.  In 
addition, we have known that spiral patterns develop spontaneously 
in $N$-body simulations ever since Lindblad's pioneering work in the 
early 1960s (\eg\ Lindblad 1960).  The problem is thus largely one 
of classical dynamics -- an application of nothing more 
sophisticated than Newton's laws of gravity and motion -- with gas 
playing an important, but secondary, role.

Spirals in tidally interacting galaxies could well result from the 
interaction itself, and some spirals may be driven by bars.  But 
spirals in a substantial fraction of galaxies cannot be ascribed to 
either of these triggers, and therefore present the most insistent 
problem.  In this article, I describe my recent work on this 
subject, which stems from my collaboration with Franz Kahn in the 
late 1980s.  He was a great help to me then, as he had been much 
earlier when I began my career as a graduate student in Manchester.  
It is clear that he had been interested in spiral structure 
throughout his career, and was present at the seminal meeting 
(Woltjer 1962) which seems to mark the change of focus in the wider 
community to gravitational theories for the phenomenon (which B. 
Lindblad had pioneered for many years before then).

\section{Short- or Long-lived Patterns?}
While most theorists agree that spirals are density variations in 
the disc which are organized by gravity, opinions diverge quite 
quickly from this starting point.  There is not even a consensus on 
something as fundamental as the lifetime of spiral patterns.  Our 
snapshot view of galaxies gives us no direct information, and two 
separate schools of thought exist on the longevity of the structures 
we observe.

C. C. Lin and his co-workers (\eg\ Bertin \& Lin 1996) favour 
long-lived quasi-stationary spiral patterns.  They suggest that 
spirals are global instabilities which grow rather slowly in a cool 
disc with a smooth distribution function (\DF).  Such modes could be 
quasi-stationary because of mild non-linear effects, such as gas 
damping.  They can achieve low growth rates for spiral modes by 
invoking a ``$Q$-barrier'' where in-going travelling waves reflect 
off a dynamically hot and largely unresponsive inner disc.  The 
standing wave pattern which makes up the mode consists of short and 
long trailing waves that are trapped between reflections at the 
$Q$-barrier and at co-rotation; it is excited by mild 
over-reflection at co-rotation.  This type of mode can survive only 
if the inner reflection occurs outside the inner Lindblad resonance 
(ILR) so that the waves are shielded from the fierce damping which 
must occur if that resonance were exposed.

Most other workers favour short-lived patterns, with fresh spirals 
appearing in rapid succession, as indicated by $N$-body simulations. 
 Such spirals develop through swing-amplification in some form or 
another, either as shearing waves (Goldreich \& Lynden-Bell 1965) or 
as forced responses (Julian \& Toomre 1966).  Both these 
interpretations are manifestations of the same underlying mechanism 
(see Toomre 1981).

The role of gas in this picture is as follows:  All forms of density 
wave arise through collective motions of the stars and are therefore 
weaker when stars move more randomly.  Thus fluctuating spiral 
structure is self-limiting, since the transient patterns themselves 
gradually scatter stars away from near-circular orbits.  If no gas 
were present, the spirals must fade over time -- in less than ten 
galactic rotations (Sellwood \& Carlberg 1984) -- as stellar random 
motions rise.  Clouds of gas, when present in a galaxy disc, collide 
and dissipate most of the random motion they acquire, and therefore 
remain a dynamically cool and responsive component.  Moreover, young 
stars reduce the rms spread of the total stellar distribution, 
because they possess similar orbits to those of their parent massive 
gas clouds.  Quite a modest star formation rate is needed to 
preserve the participation of the stellar disc in spiral waves -- a 
few solar masses per year over the disc of a galaxy is enough.

Toomre \& Kalnajs (1991) advocate one theory of this type in which 
spirals are the polarized disc response to random density 
fluctuations.  If the $\sim 10^{10}$ individual stars of the disc 
were smoothly distributed, density variations would be tiny; but 
real galaxy discs are much less smooth because they contain a number 
of massive clumps, such as star clusters and giant molecular clouds 
(an extra role for gas).  Density fluctuations can be decomposed 
into a spectrum of plane waves of every pitch angle, which shear 
continuously from the leading to the trailing direction because of 
differential rotation.  The smooth background disc responds 
enthusiastically to forcing from this shearing spectrum of density 
fluctuations, and the entire disc develops a transient spiral 
streakiness.

An equivalent viewpoint is to regard the background disc as 
polarized, with each of the orbiting disturbance masses inducing a 
spiral response in the surrounding medium.  The spiral responses are 
much stronger than the forcing density variations, but remain 
directly proportional in this picture, the proportionality constant 
depending on the responsiveness of the background disc.  The shot 
noise from the finite number of particles is itself the source of 
the spiral responses in the $N$-body simulations of Toomre \& 
Kalnajs.  While these authors have developed a rather detailed 
understanding of their local simulations, the amplitudes of spirals 
in global $N$-body simulations seems to be independent of the 
particle number (Sellwood 1989), rather than declining as $N^{-1/2}$ 
as this theory would predict.  Further, it is unclear whether the 
noise amplitude/responsiveness combination are sufficient to give 
rise to spirals of the amplitude we observe.

I prefer a model in which the spirals are true instabilities, which 
recur in rapid succession, as a flag flaps repeatedly in a breeze.  
I describe this idea in the next few sections, but I emphasize here 
that it differs radically in two further respects from the 
quasi-steady waves advocated by Bertin \& Lin.  First, the ILR in my 
model is not shielded, but plays a central role, and second, I 
expect the \DF\ to be far from smooth and, in fact, the features in 
the \DF\ drive the instabilities.  I do not claim a fully worked out 
theory, however; the weakest link in the picture is the manner in 
which the cycle recurs.

\begin{figure}[t]
\centerline{\psfig{figure=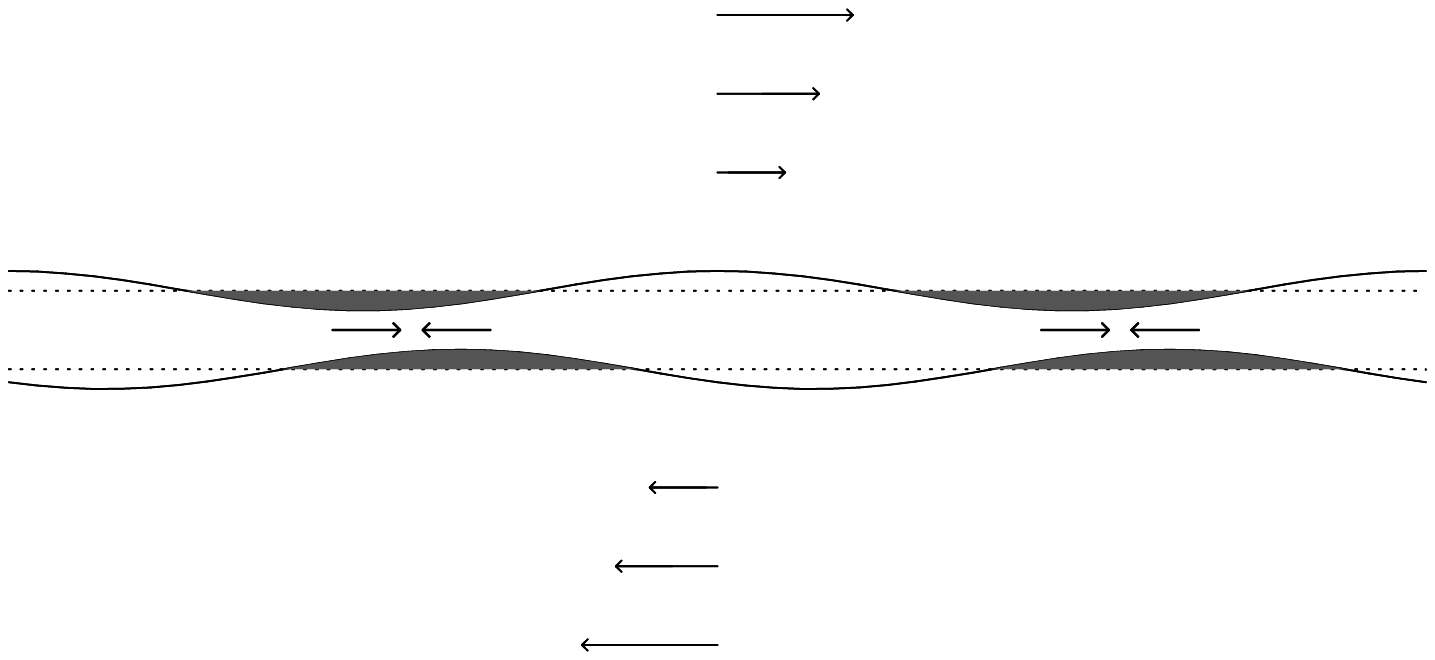,width=0.8\hsize,angle=0}}
\centerline{\psfig{figure=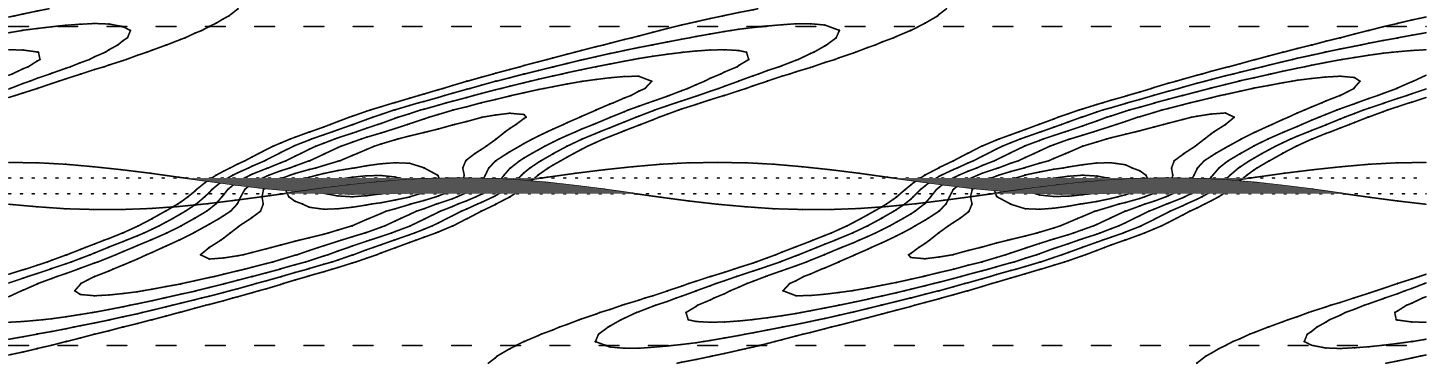,width=0.8\hsize,angle=0}}
\caption{The top panel shows the mechanism of the groove mode.  The 
surface density is lower between the two lines, which were 
originally straight (dotted) and the sense of the shear flow is 
indicated by the lightly drawn arrows.  The shaded areas are regions 
in which higher density material has moved into the groove owing to 
the disturbances on each side, and the destabilizing disturbance 
forces within the groove are marked by the heavy arrows. \hfil\break 
The density excesses in the groove are redrawn in the lower panel 
where the contours show the supporting response of the surrounding 
disc.  The dashed lines mark the Lindblad resonances.}
\label{fig:groove}
\end{figure}

\section{Groove Modes}
My paper with Franz Kahn, which appeared in {\it Monthly Notices} in 
1991, described a new kind of global spiral instability of a stellar 
disc, which we called a ``groove mode.''  Our paper presents both 
$N$-body simulations and a local theoretical description of the 
instability.  I attempt to provide only a physical interpretation 
here and leave the interested reader to refer back to our paper for 
a more thorough development. 

As its name implies, the instability occurs in a ``groove'' in the 
particle distribution.  It is fundamentally a groove in angular 
momentum density which would give rise to a groove in surface 
density only if the stars were all on circular orbits.  As the 
epicyclic radii of typical stellar orbits are some 10\% -- 20\% of 
their mean radius, whereas we invoke a deficiency of stars over a 
range of $\ltsim1$\% of their angular momentum, the surface density 
of stars is not significantly reduced anywhere.  Nevertheless, it is 
much easier to envisage the mechanism for a cold disc in which the 
radial density profile has a sharp notch.

The mechanism for a groove mode is illustrated in Figure 
\ref{fig:groove} which shows a small patch of the disc so far from 
its centre that curvature is negligible.  Wave-like disturbances on 
the groove edges bring high surface density material into the groove 
in the dark shaded regions.  The changes in density give rise to 
disturbance forces; those between the density excesses on either 
side of the groove are marked by the pairs of opposing arrows within 
the groove.  Material displaced upwards into the groove from its 
lower edge is therefore pulled forward by the density excess from 
the opposite edge.  Material that is pulled forward gains angular 
momentum, moves to an orbit of larger mean radius, and in a 
differentially rotating disc, lags behind its original azimuthal 
motion.  It therefore rises further into the groove, moving less 
rapidly, relative to the groove centre than before.  Similarly, 
material displaced downwards into the groove from its upper edge is 
pulled backward by the density excess from the opposite edge, loses 
angular momentum and sinks further into the groove.  Thus each 
density excess causes the other to continue to grow, and the system 
is unstable.

It is worth mentioning that Franz Kahn did not see the instability 
this way.  Instead, he instinctively saw that the dispersion 
relation for waves in a groove with a smooth profile, such as a 
Lorentzian, would have a solution in the upper half of the complex 
plane.  To my mind, such an insight conveys no intuitive feel for 
what is happening, but Franz rightfully trusted it more than 
physical intuition in situations where the dynamics can be quite 
subtle.

The instability of the groove alone would be of little consequence 
were it not for the supporting response of the surrounding disc.  A 
polarized response grows with the density excesses in the groove, as 
shown in the lower part of Figure \ref{fig:groove}.  The groove 
itself is unstable over a wide range of wavelengths, but the 
vigorous supporting response from the surrounding disc favours long 
wavelengths only.  The result is a fiercely growing, global, spiral 
instability for which co-rotation lies near the groove centre.  Our 
local theory estimates of the mode frequencies based on this picture 
(Sellwood \& Kahn 1991) were in tolerable agreement with those found 
in the global simulations.

These spiral responses extend as far as the Lindblad resonances on 
either side, which are marked by the dashed lines.  Recall that 
these resonances occur where stars moving relative to the pattern 
encounter the periodic disturbance at their epicyclic frequency, 
$\kappa$.  If the pattern speed is $\Omega_p$ and the circular 
frequency at radius $R$ is $\Omega(R)$, a star encounters an 
$m$-armed wave at frequency $\omega=m(\Omega-\Omega_p)$; the 
condition for a Lindblad resonance is $|\omega| = \kappa$.

I should stress that this linear, global instability also occurs in 
discs with random motion, in which case we invoke a deficiency of 
stars over a narrow angular momentum range.  The above arguments 
carry over to this more realistic situation.

\section{Subsequent behaviour}
The linear mode saturates, and ceases to grow, about the time that 
the density excesses meet across the groove.  At this point, the 
mode has generated a periodic density variation around the groove 
which will disperse only rather slowly, as the stars which comprise 
it all have similar angular momenta.  The amplitude of the polarized 
response of the surrounding disc tracks the variations of the mass 
clump which induced it.

\subsection{Angular momentum transport}
The inclined spirals exert a torque, however, which redistributes 
angular momentum.  The effect can be viewed as a gravitational 
stress (Lynden-Bell \& Kalnajs 1972), or as wave action carried at 
the group velocity (Toomre 1969; Kalnajs 1971).  The group velocity, 
$\partial \omega / \partial k$ as usual, is directed radially away 
from co-rotation for all but the most open trailing waves (Binney \& 
Tremaine 1987, \S6.2).  As the wave is being set up, the stars 
inside co-rotation do work on the wave and lose angular momentum to 
it, while those outside gain energy and angular momentum; the net 
change over the whole disc is zero, as it must be for a self-excited 
disturbance.  Thus the wave action carried at the group velocity is 
negative angular momentum inside co-rotation and positive angular 
momentum outside.  When combined with the opposite signs of the 
group velocity, there is an outward flux of angular momentum 
everywhere -- in agreement with the sign of the gravity torque.

\subsection{Exchanges at resonances}
The disturbance produced by the groove mode therefore generates both 
positive and negative angular momentum, in equal measure, at 
co-rotation which is then carried away by the spirals towards the 
Lindblad resonances on either side.  Secular exchanges between the 
wave and the stars, which are possible only at resonances 
(Lynden-Bell \& Kalnajs 1972), lead to the wave action being 
absorbed at these locations, which is why a mode could not be set up 
with an exposed ILR.

\begin{figure}[t]
\centerline{\psfig{figure=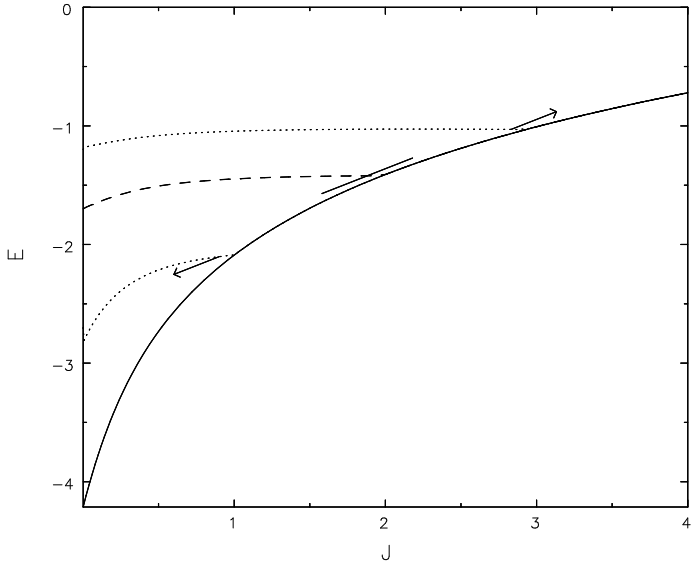,width=0.7\hsize,angle=0}}
\caption{The Lindblad diagram for a disc galaxy model.  Circular 
orbits lie along the full-drawn curve and eccentric orbits fill the 
region above it.  Angular momentum and energy changes between a wave 
and particles move them along lines of slope $\Omega_p$ as shown.  
See the text for explanations of the other curves.}
\label{fig:lindblad}
\end{figure}

Stars moving in a non-axisymmetric potential that rotates at a 
steady rate conserve neither their specific energy, $E$, nor their 
specific angular momentum, $J$.  But the combination
\begin{equation}
I_{\rm J} \equiv E - \Omega_p J,
\end{equation}
known as Jacobi's invariant, is conserved.  Thus at a Lindblad 
resonance, where a star gains angular momentum $\Delta J$ from the 
wave, it also changes its energy as
\begin{equation}
\Delta E = \Omega_p \Delta J.
\end{equation}

Figure \ref{fig:lindblad} shows the Lindblad diagram for a 
differentially rotating disc with an infinitesimal non-axisymmetric 
perturbation.  The full-drawn curve marks the locus of circular 
orbits in this $(J,E)$ plane; no particle can lie below this curve, 
but bound particles with $E>E_c$ move on eccentric orbits in this 
potential.  The resonance condition for a non-circular orbit 
generalizes to 
\begin{equation}
m(\Omega_\phi - \Omega_p ) + l \Omega_R = 0,
\end{equation}
where $\Omega_\phi(E,J)$ and $\Omega_R(E,J)$ are the azimuthal and 
radial frequencies of orbits (Binney \& Tremaine 1987, \S3.1), and 
$l=0$ for co-rotation and $l=\pm1$ for Lindblad resonances.  The 
loci of these three principal resonances for arbitrarily eccentric 
orbits are marked by the broken curves in this Figure.

The displacements caused by wave-particle interactions all have 
slope $\Omega_p$ in this diagram.  As this is the slope of the 
circular orbit curve at co-rotation, stars which exchange energy and 
angular momentum there do not move further from that curve, to first 
order.  The two vectors show that at the Lindblad resonances, on the 
other hand, stars are moved onto more eccentric orbits when angular 
momentum is redistributed outwards.

\begin{figure}[t]
\centerline{\psfig{figure=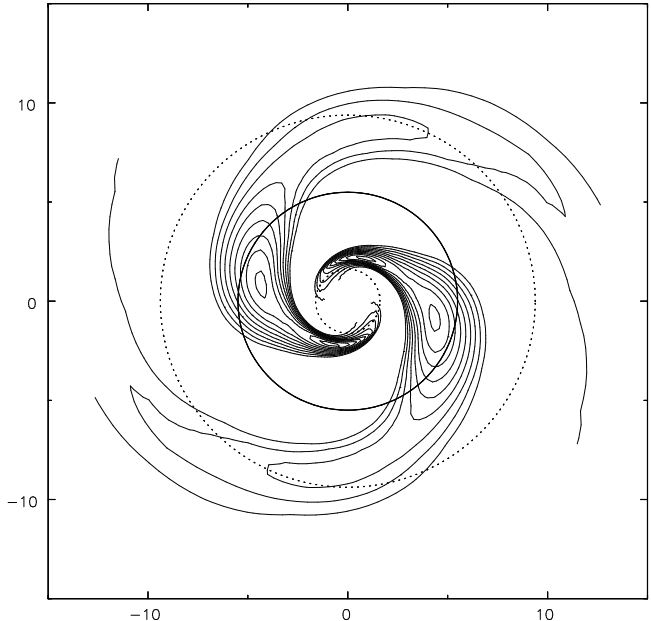,width=0.65\hsize,angle=0}}
\caption{The best fit coherent wave fitted to the $m=2$ data from 
the simulation described in \S4.3.  The full-drawn circle marks the 
radius of co-rotation while the dashed circles indicate the Lindblad 
resonances for this pattern.}
\label{fig:mode}
\end{figure}

\subsection{Results from a simulation}
Figure \ref{fig:mode} shows a spiral pattern extracted from one of 
my simulations.  This calculation is of a $Q=1.5$, half-mass 
$V=V_0=\;$const (aka Mestel) disc, with a inner taper applied to 
give the disc a characteristic length scale, $R_0$, and an outer cut 
off to limit it to a finite radial extent.  The properties of this 
disc are described in Binney \& Tremaine (1987, \S4.5) and I have 
previously reported (Sellwood 1991) confirmation of Toomre's (1981) 
prediction that a this model with a smooth \DF\ has no true linear 
instabilities.  The 1 million particles in this simulation were 
drawn from the appropriate \DF\ and placed at random azimuths at the 
start.  A number of transient spiral patterns develop from the 
particle noise over time, and that illustrated in Figure 
\ref{fig:mode} was extracted by fitting a coherent wave to $m=2$ 
Fourier coefficients of the mass distribution for the time interval 
$400 \leq t V_0/R_0 \leq 600$ -- \ie\ long after the start.  The 
best-fit pattern speed for this wave is $2\Omega_p = 0.364V_0/R_0$ 
and co-rotation (full drawn circle) and the Lindblad resonances 
(dotted circles) for this pattern speed are marked.

\begin{figure}[p]
\centerline{\psfig{figure=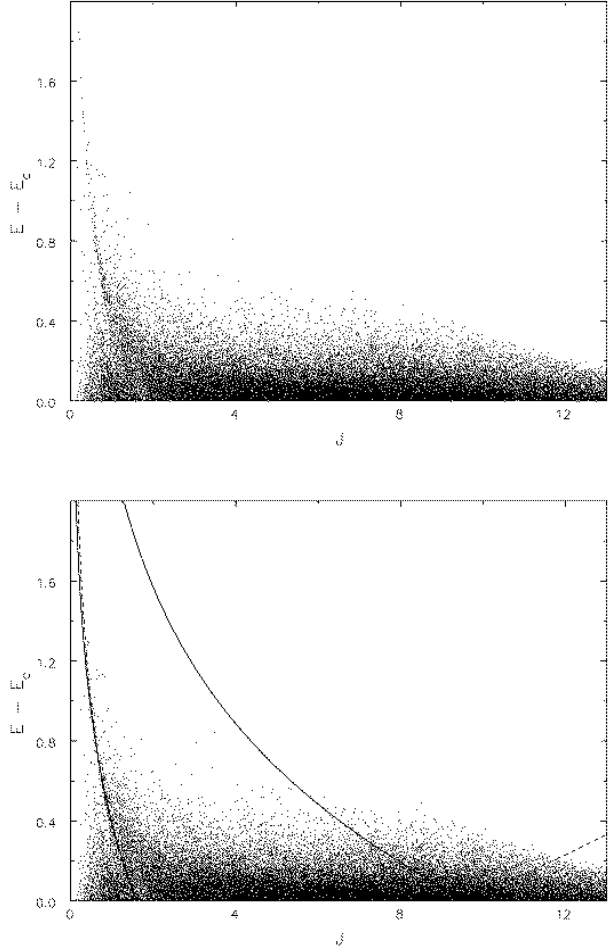,width=0.9\hsize,angle=0}}
\caption{Top: The locations of one particle in 20 in the simulation 
in $(E_{\rm ran}, J)$-space at a time during the decay of the 
pattern shown in Figure \ref{fig:mode}.  Bottom: The same with added 
curves to show (full-drawn the loci of the generalized Lindblad 
resonances (equation 3) and (dashed) to mark the scattering 
trajectory given by equation (2) for particles which start with 
$E=E_c$.}
\label{fig:phasesp}
\end{figure}

The distribution of particles in this run at time 600 is shown in 
Figure \ref{fig:phasesp}.  The abscissae are the (instantaneous) 
specific angular momenta of the particles, and the ordinates are the 
excess of energy over and above that needed for a circular orbit at 
this $J$.  Overlayed on the lower plot are the expected loci of 
particles scattered from nearly circular orbits at the Lindblad 
resonances for this wave, and the loci of the generalized Lindblad 
resonances defined by equation (3).  The very close similarity in 
this potential between the scattering trajectory and the locus of 
the generalized ILR is remarkable, and leads to the strong and 
coherent tongue of particles scattered up this curve.   The 
impressive agreement between the simulation and the prediction, 
which has no free parameters, is reassuring.

The near coincidence of the scattering trajectory with the 
generalized ILR is not repeated at the OLR; there the two curves are 
practically orthogonal.  Consequently, scattering at the OLR does 
not produce striking changes in the particle distribution in this 
plot.

\subsection{Recurrent cycle}
The resonant scattering of stars by spiral waves, leading to 
deficiencies in the \DF\ over narrow regions in this plot, opens up 
the possibility of a recurrent cycle.  Similar, though not 
identical, deficiencies are responsible for the groove modes 
discovered by Franz and myself, and it is therefore likely that a 
fresh instability should develop, perhaps with co-rotation near to 
one of the Lindblad resonances.

A recurrent instability cycle of this nature was observed by 
Sellwood \& Lin (1989) in simulations of a low mass disc in nearly 
Keplerian rotation about a central mass.  Their model, in which 
perturbation forces were also restricted to a single Fourier 
component, differs in many important respects from real galaxies, 
but the present Mestel disc is much more realistic.  Scattering 
caused by the waves raises the level of random motion of disc 
particles in the vicinity of the resonance.  Sellwood \& Lin found 
that only the groove carved in a previously undisturbed part of the 
disc caused a new mode to grow -- the other Lindblad resonance lay 
in a region which had already been heated by two previous waves, and 
was unable to support a new instability.  Thus the spiral 
``disease'' progresses radially in a single direction.  Once the 
entire disc has been heated, no further spiral waves can be 
sustained -- unless some cooling mechanism is in effect (\eg\ 
Sellwood \& Carlberg 1984; Carlberg \& Freedman 1986).

\section{Hipparcos stars}

This model for a recurrent spiral wave instability cycle is now 
rather complex, and rests heavily on results from simulations.  
While the simulations have been conducted with as much care as 
possible, and their behaviour seems physically reasonable, the 
possibility always exists that the results arise through some 
pernicious, undetected numerical artifact and bear no relation to 
what actually happens in real galaxies.  It therefore seemed 
appropriate to me to seek some observational confirmation before 
plunging any further down what could be a blind alley.

A possibility of an observational test presented itself once the 
data became available from ESA's Hipparcos satellite.  This space 
mission measured proper motions and parallaxes for many stars.  Of 
these, some 14\,000 were selected by Binney \& Dehnen (1998) as 
being a kinematically unbiased subsample within 100 pc of the Sun, 
so that parallaxes were known to a precision of 10\% or better.  The 
satellite measured five out of the six phase-space coordinates of 
each of these stars -- the radial velocity was not measured.  
However, Dehnen (1998) cleverly realized that the missing 
information could be obtained, at least in a statistical sense, if 
it could be assumed that the velocity distribution of these stars 
were homogeneous throughout this small volume.  This reasonable 
assumption allows one to infer the distribution of missing 
velocities in one direction on the sky by equating it to the 
distributions observed in orthogonal directions.  In this way, 
Dehnen was able to construct the full phase space distribution 
function.

\begin{figure}[t]
\centerline{\psfig{figure=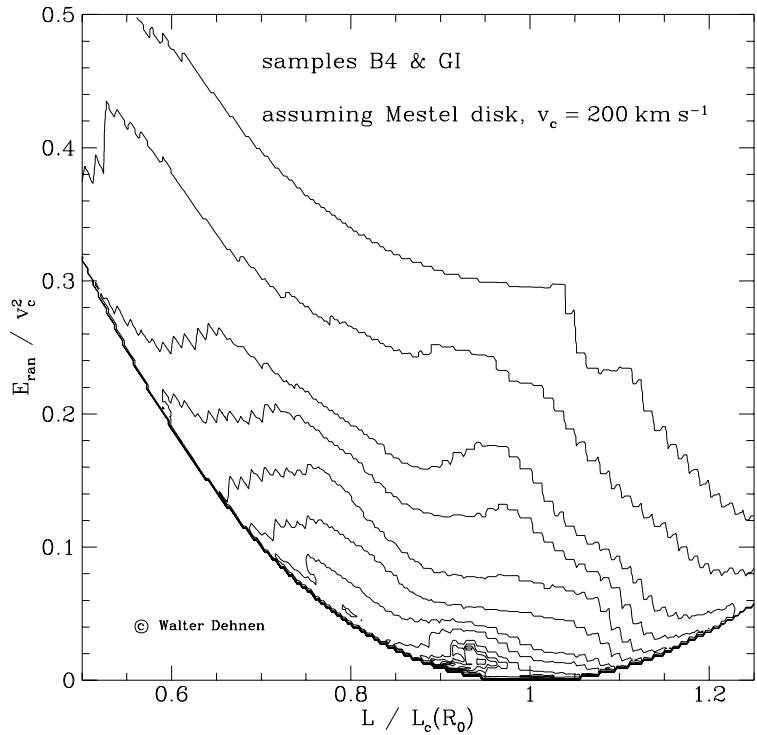,width=0.8\hsize,angle=0}}
\caption{The density of Solar neighbourhood stars in $(E_{\rm 
ran},L)$-space.  The figure was constructed by Dehnen using $\sim 
6\,000$ stars selected from the Hipparcos sample.  Apart from the 
skew to lower $L$, which is caused by the asymmetric drift, a smooth 
\DF\ would have produced a featureless plot, whereas the contours 
show one or more distinct ridges.}
\label{fig:dehnen}
\end{figure}

Figure \ref{fig:dehnen}, kindly made for me by Dehnen, shows 
contours of the phase space density of Solar neighbourhood stars in 
$(E_{\rm ran}, J)$-space.  The parabolic lower boundary of the 
contour distribution reflects the fact that stars on close to 
circular orbits, whose guiding centers are far from the Sun would 
never visit the solar neighbourhood, and are therefore missing from 
this sample.  The asymmetry between left and right results from the 
usual asymmetric drift, because the stellar density rises towards 
the Galactic centre.

But it is clear from this Figure that the underlying distribution 
function is not smooth.  There is a clear hint of a scattering line 
(maybe two), which is reminiscent of scattering at an ILR.  If this 
interpretation is correct, it would confirm that spiral arms are 
transient and lend considerable support to the idea of a recurrent 
transient cycle of instabilities discussed above.  The missing 
radial velocity information is now being obtained (Pont \etal\ 
1999), which will enable this plot to be redrawn without invoking 
Dehnen's stratagem and hopefully, confirm the sub-structure.

It should be noted that Dehnen (1999) suggests these data could also 
be interpreted as scattering at the OLR of the bar (see also Raboud 
\etal\ 1998).  We are working to try to determine which 
interpretation is the more plausible, but either way, the assumption 
that spiral arms in galaxies are formed in a system having a smooth 
\DF\ looks rather too idealized in the light of these data.

\section{Conclusions}
Theoretical effort in an area starved of observational input often 
loses momentum and may be aimed in quite the wrong direction.  
Spiral structure theory has not been devoid of observational input, 
since evidence in favour of density waves has been accumulating for 
many years.  However, the near IR intensity variations (Schweizer 
1976 and others), or coherent velocity perturbations (Visser 1978 
and others), support only the existence of density waves, and do not 
test ideas for their origin.

Simulations of disc galaxies were strongly motivated by the desire 
to fill this gap, and have been reporting for decades that spiral 
arms are transient.  As this result has still not been fully 
understood, the simulators themselves have worried that it could be 
an artifact.  Such healthy scepticism has prompted them to devote 
many years of effort to refining, testing and cross-checking their 
codes (Miller 1976; Sellwood 1983; Inagaki \etal\ 1984; \etc).  But 
new results from simulations cannot, almost by definition, be 
checked, and the best we can do is to try to show that the behaviour 
is physically reasonable.  Despite all this care and effort, those 
intent on calculating slowly growing, quasi-stationary, spiral modes 
have totally discounted the reported behaviour on the grounds that, 
in their view, spiral structure is simply too ``delicate'' a problem 
for simulated results to have any validity!

The Hipparcos data provide the first observational confirmation that 
it is {\it wrong\/} to assume that the \DF\ of a disc galaxy is 
smooth.  In retrospect, it is hard to see how it could remain 
smooth, since almost any realistic disturbance in a disc will 
scatter stars from (or maybe trap them into) resonant regions of 
phase space.  Obviously, the \DF\ could relax back to a smooth state 
if scattering were efficient, but in a purely collisionless disc, 
the feature could be smoothed only by further collective effects, 
which themselves will have resonances elsewhere.

The Hipparcos result appears to vindicate the simulations, and it 
seems highly likely that spirals in real galaxies are recurring, 
transient patterns.  They result from true instabilities provoked by 
narrow features in the \DF, and are of a global nature because long 
wavelength disturbances are supported most vigorously by the 
swing-amplifier.  While the mechanism for these linear instabilities 
is now reasonably clear, exactly how the \DF\ is affected, and how 
the instabilities might recur is not.  The resonant scattering 
peaks, if confirmed, suggest at least one of the processes which 
sculptures the \DF, but it is possible it is not the only, or even 
the dominant, source of inhomogeneities in the \DF.  The Hipparcos 
data have provided a much needed pointer to the way forward in this 
erstwhile stalled area.

\acknowledgments
This work was supported by NSF grant AST 96/17088 and NASA LTSA 
grant NAG 
5-6037.

\end{document}